\documentclass[12pt]{revtex4-1}
\usepackage{amsbsy}
\usepackage{amssymb}
\usepackage{amsmath}
\usepackage{geometry}
\usepackage[british]{babel}
\usepackage{amscd}
\usepackage{fancyhdr}
\usepackage{hyperref}
\usepackage{mathrsfs}
\usepackage[cmtip,arrow]{xy}	
\usepackage{pb-diagram,pb-xy}   
\usepackage{yfonts}

	\newcommand{\ncd}{\newcommand}
	\ncd{\mrm}    {\mathrm}
	\ncd{\beq} {\begin{equation}}
	\ncd{\eeq} {\end{equation}}

	\def\d{{\rm d}}

\begin{document}

	\title{Comment on ``Geometrothermodynamics of a Charged Black Hole of String Theory''}

	\author{C. S. Lopez-Monsalvo}
	\affiliation{Instituto de Ciencias Nucleares\\
     			 Universidad Nacional Aut\'onoma de M\'exico, A. P. 70-543, M\'exico D. F. 04510, M\'exico}
	
	\author{F. Nettel}
	\affiliation{Departamento de F\'\i sica, Fac. de Ciencias \\
			 Universidad Nacional Aut\'onoma de M\'exico, A. P. 50-542, M\'exico  D. F. 04510, M\'exico}
	
	\author{A. S\'anchez}
	\affiliation{Departamento de Posgrado, CIIDET, A. P. 752, Quer\'etaro, Qro 76000, M\'exico}

	\date{\today}

	\begin{abstract}
	We comment on the conclusions found by Larra\~naga and Mojica \cite{brazilGTD} regarding the consistency of the Geoemtrothermodynamics programme to describe the critical behaviour of a  Gibbons-Maeda-Garfinkle-Horowitz-Strominger charged black hole. We argue that making the appropriate choice of metric for the thermodynamic phase space and, most importantly, considering the homogeneity of the thermodynamic potential we obtain consistent results for such a black hole. 
	\end{abstract}

\maketitle

\section{Introduction}

The recent efforts to tackle the thermodynamics of black holes from a geometric perspective have lead us to some interesting developments and results. In particular, it has become a useful tool to analyse the critical behaviour of such systems \cite{Rupp1, Rupp2, Aman, Unifiedgtd, phasetransGTD}. 

In this letter, we address some issues raised  by Larra\~naga and Mojica \cite{brazilGTD}. In their article, they  concluded that the Geometrothermodynamics programme (GTD) \cite{quevedo.gtd} does not reproduce consistently the thermodynamics of the Gibbons-Maeda-Garfinkle-Horowitz-Strominger (GMGHS)  charged black-hole \cite{GMGHS}. We argue that such a conclusion is erroneous due to an incorrect implementation of the geometric formalism.

In the following discussion we denote the thermodynamic phase space by $\mathcal{T}$, the space of equilibrium states by $\mathcal{E}$ and use the letters $G$ and $g$ for their associated metrics, respectively. The notation for the thermodynamic variables is clear from the context. Thus, we use $S$ to denote the entropy, $M$ the  mass, $T$ represents the Hawking temperature, $Q$ is the charge and $\phi$ the electromagnetic potential.

\section{Notes on GTD}

The authors of \cite{brazilGTD} start their analysis by recalling the basic construction of GTD. Namely, the local expression for the phase space contact 1-form 
	\beq
	\label{oneform}
	\Theta = \d \Phi - \delta_{ab} I^a \d E^b,
	\eeq
together with the Legendre embedding, $\varphi:\mathcal{E} \longrightarrow \mathcal{T}$, defining the space of equilibrium states. They correctly point out that the thermodynamic potential $\Phi$ must satisfy the \emph{homogeneity condition}
	\beq
	\label{homo}
	\Phi(\lambda E^a) = \lambda^\beta \Phi(E^a),
	\eeq
for some constant parameters $\lambda$ and $\beta$. In such a case we say that the thermodynamic potential is homogeneous of order $\beta$. This is a subtle point which one should consider carefully, as we argue below [c.f. equations \eqref{gtd.srep} and \eqref{gtd.mrep}]. 

 The authors also elaborate on a crucial requirement in the GTD formalism, the Legendre invariance of the metric structure $G$ of $\mathcal{T}$, which is inherited by the induced metric $g = \varphi^*(G)$ on $\mathcal{E}$.  Larra\~naga and Mojica are aware that there  is a vast number of metrics for $\mathcal{T}$ which comply to this requirement. Should we found that the metric we use leads to inconsistencies, we can only conclude that our choice was an unfortunate one. The selection rule for the thermodynamic metric remains an open issue in GTD. Nevertheless, there are sound indications \cite{phasetransGTD} that phase transitions of first and second order are reproduced by two different metrics for phase space $\mathcal{T}$. In particular for the geometric description of the thermodynamics of black holes the phase space metric is given by
	\beq
	\label{ps.G}
	G = \Theta^2 + \left(\delta_{ab} I^a E^b\right)\left(\eta_{cd} \d I^c \d E^d\right),
	\eeq 
where $\eta = \rm{diag}(-1,1,...,1)$.  Indeed, this choice reproduces the correct thermodynamic behaviour of the GMGHS -- and many other \cite{phasetransGTD} -- black-hole, as we will shortly show. 

The election for the metric $G$ for $\mathcal{T}$ [equation (7)] made by Larra\~naga and Mojica describes first order phase transitions, but fails  when it comes down to second order transitions. According to Davies \cite{Davies} phase transitions in black holes are of the second order. Although it is not clear if the extremal limit of black holes constitutes a phase transition to naked singularities, there is some indication that in any case they would be of second order \cite{Cai}. Then, the metric chosen in \cite{brazilGTD} is expected to yield misleading conclusions. Let us note that the induced metric $g$  found in \cite{brazilGTD} does not correspond to the pullback of the phase space metric $G$. Their expression [equation (8)] is only valid when the thermodynamic potential is \emph{homogeneous of order one}. This is relevant because in their subsequent analysis the metric used for the space of equilibrium states is not the pullback of the phase space metric for the fundamental relations they consider [c.f. equations \eqref{gtd.srep} and \eqref{gtd.mrep}, below].  

Choosing the suitable metric for describing black holes (\ref{ps.G}) the induced metric on the space of equilibrium states $\mathcal{E}$ is
\beq
	\label{eq.g}
	g = \left(E^f \frac{\partial \Phi}{\partial E^f}\right) \eta_{ab}\delta^{bc}\frac{\partial^2 \Phi}{\partial E^c \partial E^d} dE^a dE^d.
	\eeq

\section{GTD of the GMGHS charged black-hole}

The analysis performed by the authors considers two different (although equivalent) thermodynamic representations. On the one hand, they use the fundamental equation for the entropy obtained through the Hawking relation \cite{GMGHS} 
	\beq
	\label{gtd.srep}
	S(M,Q) = 4 \pi M^2 - 2 \pi Q^2 e^{-2 \psi_0},
	\eeq
where $e^{-2 \Psi_0}$ represents a dilaton filed. From the homogeneity condition, equation \eqref{homo}, it is clear that the entropy is a homogeneous function of order 2. Thus, one would have to use the general form \eqref{eq.g} for the induced metric. Let us make the whole calculation instead. The phase space metric \eqref{ps.G} expressed in this representation takes the form
	\beq
	G_S = \left(\d S - \frac{1}{T} \d M + \frac{\phi}{T}\d Q\right)^2 + \left(\frac{M}{T} - \frac{\phi}{T}  Q\right)\left[-\d\left(\frac{1}{T}\right)\d M  +  \d\left(-\frac{\phi}{T}\right) \d Q  \right],
	\eeq
and the induced metric, $g_S = \varphi^*(G_S)$,  is
	\beq
	g_S = -32 \pi^2 \left( 2M^2 - Q^2 e^{-2\psi_0}\right) \left[\d M^2 + \frac{1}{2}e^{-2\psi} \d Q^2 \right].
	\eeq
It is a direct calculation to see that the scalar curvature in this representation is not identically zero, as found in \cite{brazilGTD}, but instead is given by the richer expression
	\beq
	\label{Rgtd.s}
	R_S = -\frac{1}{4\pi^2}\left[\frac{2 M^2 + Q^2 e^{-2\psi_0}}{2 M^2 - Q^2 e^{-2\psi_0}}\right].
	\eeq
Note that a curvature singularity occurs when 
	\beq
	\label{naked}
	M^2 = \frac{1}{2} Q^2 e^{-2\psi_0},
	\eeq
Such behaviour is indicative of a phase transition \cite{quevedo.2008}. Physically, equation \eqref{naked} represents the appearance of a naked singularity \cite{GMGHS}. 

On the other hand, they also make the analysis using the mass representation, i.e.
	\beq
	\label{gtd.mrep}
	M(S,Q) = \sqrt{\frac{S}{4 \pi} + \frac{1}{2}Q^2 e^{-2\psi_0}}.
	\eeq
Now the situation is harder to describe, since the mass is not a homogeneous function of the entropy and charge. Nevertheless, we can use the standard procedure of GTD. In this case, the phase space metric is written as 
	\beq
	G_M = \left(\d M - T \d S - \phi \d Q\right)^2 + \left(TS + \phi Q\right)\left[-\d T \d S + \d \phi \d Q\right],
	\eeq
and its pullback to the space of equilibrium states is given by
	\beq
	g_M = \frac{1}{4\pi} \left[\frac{S + 4 \pi Q^2 e^{-2\psi_0}}{\left(S + 2 \pi Q^2 e^{-2\psi_0}\right)^2}\right]\left(\frac{1}{8} \d S^2 + \pi  e^{-2\psi_0} S \, \d Q^2\right).
	\eeq
The scalar curvature takes the form
	\beq
	\label{RM}
	R_M = 64 \pi^2Q^2 e^{-2\psi_0}\left[\frac{(S + 2 \pi Q^2 e^{-2\psi_0} )^2 \, (3S +4 \pi Q^2 e^{-2\psi_0})}{(S + 4\pi Q^2 e^{-2\psi_0})^3S^2}\right].
	\eeq
It is clear that a singularity only occurs when the entropy is identically zero. This indicates the occurrence of a phase transition, consistent with the result using the entropy representation [c.f. equations \eqref{gtd.srep} and \eqref{naked}].

\section{Closing remarks}

We have carried out the thermodynamic analysis of the GMGHS black hole using the GTD formalism. Let us note that we have made a different choice for the phase space metric according to the analysis given by Quevedo et.al. \cite{phasetransGTD}. This choice has given consistent results with other black hole solutions and, as follows from equations \eqref{Rgtd.s} and \eqref{RM}, we observe the same curvature structure for both, entropy and mass, representations. Moreover,  Larra\~naga and Mojica do not use the correct expression for the induced metric in the space of equilibrium states [c.f. equation \eqref{eq.g}]. Instead, they use a form which is valid only when the thermodynamic potential is a homogeneous function of order one in the extensive variables, which is not the case for \eqref{gtd.srep} and \eqref{gtd.mrep}. 

Let us finally note that the crucial feature of the GTD programme is the invariance of the thermodynamic description under change of coordinates (Legendre transformations) or potentials. The disagreement in the curvature scalar in two different representations is always a strong indication that the metric choice is inappropriate.

The complete thermodynamic analysis of the GMGHS solution will appear shortly in a forthcoming paper \cite{inprep}.

\section*{Acknowledgements}

The authors are thankful to H. Quevedo for insightful comments and discussions. This work was partially supported by DGAPA-UNAM No. IN106110 grant and CONACYT-Mexico, grant No. 166391.  F. N. acknowledges support from DGAPA-UNAM (Post-doctoral Fellowship).

\end{document}